\documentclass{reportOpenlab} % custom style based on LaTeX article
\begin{document}

%--------------------- Define project ------------------------
\title{AI-Enhanced Operator Assistance for UNICOS Applications}
\author{Bernard Tam}
\institute{The University of Sydney}
\supervisor{Jean-Charles Tournier\\Fernando Varela Rodriguez}
\date{August 2025}
\reportnumber{2025}

%--------------------- Produce first pages ------------------------
\maketitlepage

\abstract{\noindent This project explores the development of an AI-enhanced operator assistant for UNICOS, CERN’s UNified Industrial Control System. While powerful, UNICOS presents a number of challenges, including the cognitive burden of decoding widgets, manual effort required for root cause analysis, and difficulties maintainers face in tracing datapoint elements (DPEs) across a complex codebase. In situations where timely responses are critical, these challenges can increase cognitive load and slow down diagnostics.\\

\noindent To address these issues, a multi-agent system was designed and implemented. The solution is supported by a modular architecture comprising a UNICOS-side extension written in CTRL code, a Python-based multi-agent system deployed on a virtual machine, and a vector database storing both operator documentation and widget animation code. Three specialised agents were developed to work under a single supervisor agent: one to automate widget decoding, one to perform root cause analysis by traversing device hierarchies, and one to trace the DPEs responsible for widget animation.\\

\noindent Preliminary evaluations suggest that the system is capable of decoding widgets, performing root cause analysis by leveraging live device data and documentation, and tracing DPEs across a complex codebase. Together, these capabilities reduce the manual workload of operators and maintainers, enhance situational awareness in operations, and accelerate responses to alarms and anomalies. Beyond these immediate gains, this work highlights the potential of introducing multi-modal reasoning and retrieval augmented generation (RAG) into the domain of industrial control.\\

\noindent Ultimately, this work represents more than a proof of concept: it provides a basis for advancing intelligent operator interfaces at CERN. By combining modular design, extensibility, and practical AI integration, this project not only alleviates current operator pain points but also points toward broader opportunities for assistive AI in accelerator operations.}

\tableofcontents

%--------------------- Include main content ------------------------

\section{BACKGROUND}

\subsection{UNICOS}
UNified Industrial Control System (UNICOS) is a framework at CERN to build industrial control systems \cite{gayet2005unicos}. More than 200 systems at CERN run on UNICOS, including the technical infrastructure for the accelerator complex \cite{goralczyk2022cern}. UNICOS development is carried out using WinCC OA, a Supervisory Control and Data Acquisition (SCADA) system developed by Siemens to monitor and control automated processes. Programs in WinCC OA are written in CTRL code \cite{silvola2022devops}, which is similar in syntax to C \cite{introductiontoctrl}. As a result, development work on UNICOS in this project will be carried out in CTRL code.

\begin{figure}[H]
\centering
\includegraphics[width=0.5\textwidth]{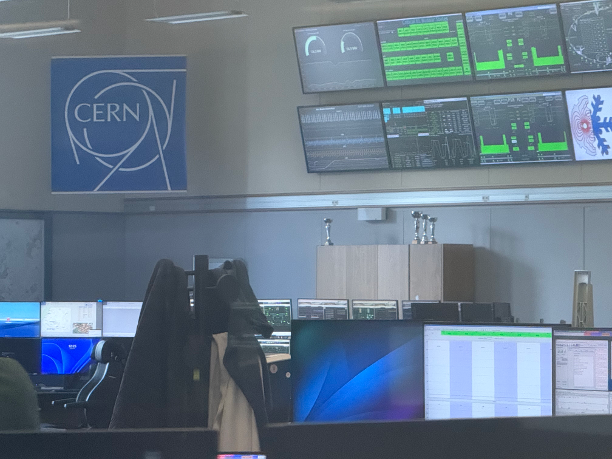}
\caption{An image of the interior of the CERN Control Centre on CERN's Prévessin site}
\label{fig:unicos_ccc}
\end{figure}

\begin{figure}[H]
\centering
\includegraphics[width=0.6\textwidth]{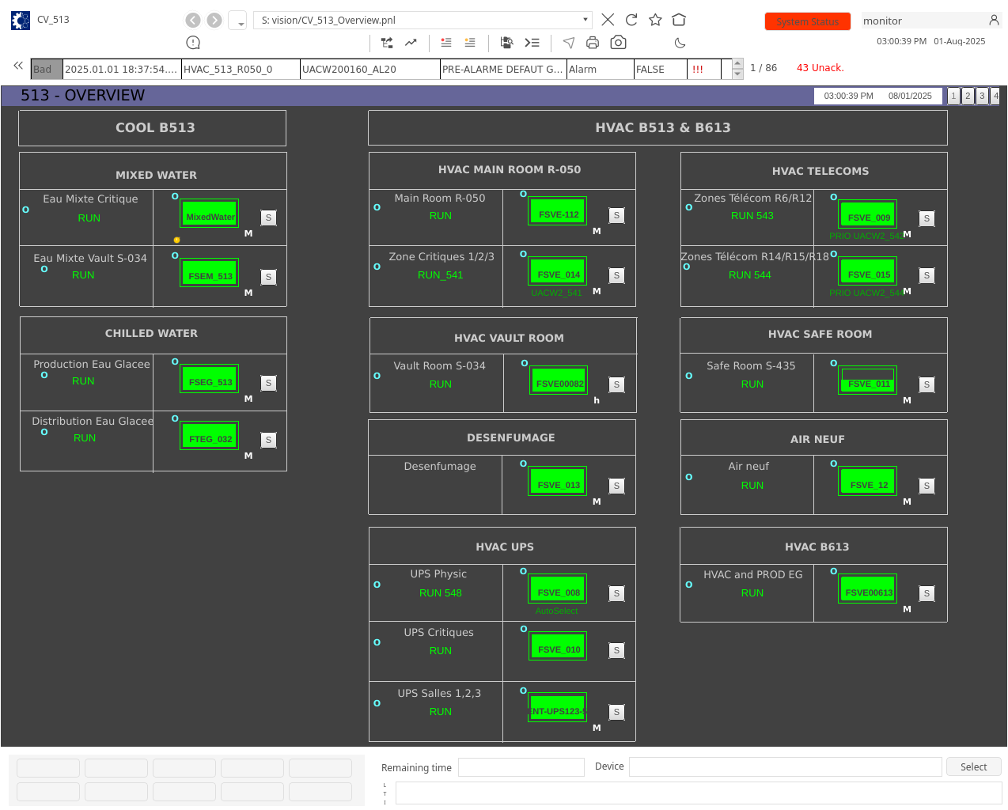}
\caption{A screenshot showing an example of a UNICOS panel}
\label{fig:unicos_panel}
\end{figure}

\noindent Figure~\ref{fig:unicos_ccc} above shows the interior of the CERN Control Centre (CCC), with the monitors in the background seen to be running UNICOS. Figure~\ref{fig:unicos_panel} below provides an example of a UNICOS panel, offering a clearer view of what is usually displayed on those monitors.

\subsection{UNICOS Pain Points and AI Solutions}
Various pain points relating to UNICOS have been experienced by different types of CERN personnel, including UNICOS operators and maintainers \cite{projectdescription}. The objective of this section is to explain each of the pain points in detail, describe their potential implications, and outline the solutions that have been developed in this work to address each pain point.\\

\begin{figure}[H]
\centering
\includegraphics[width=0.3\textwidth]{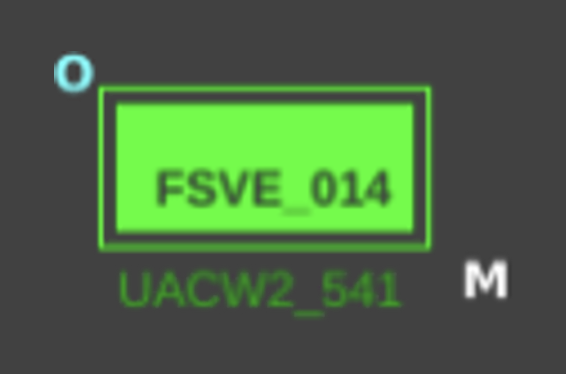}
\caption{A screenshot showing an example of a UNICOS widget}
\label{fig:unicos_widget}
\end{figure}

\noindent The first pain point for UNICOS operators manifests in the decoding of widgets. Figure~\ref{fig:unicos_widget} above shows an example of a widget in UNICOS. Correspondingly, the documentation provided to UNICOS operators that explains the symbols on each widget is displayed in Figure~\ref{fig:unicos_widget_documentation} below. When decoding a widget or multiple widgets simultaneously, operators may have to reference the documentation frequently, potentially leading to high cognitive load and information overload. This would especially be the case if the widget in question contains an uncommon symbol, thereby requiring manual reference to the documentation. This issue is exacerbated in situations where multiple widgets are indicating a critical alarm, meaning that the full details of widgets must be decoded immediately and subsequent action taken to ensure a continued safe operation of systems. This work seeks to automate the widget decoding process with the help of AI, thereby reducing the cognitive load on UNICOS operators and contributing to a continued safe operation of systems.\\

\begin{figure}[H]
\centering
\includegraphics[width=0.8\textwidth]{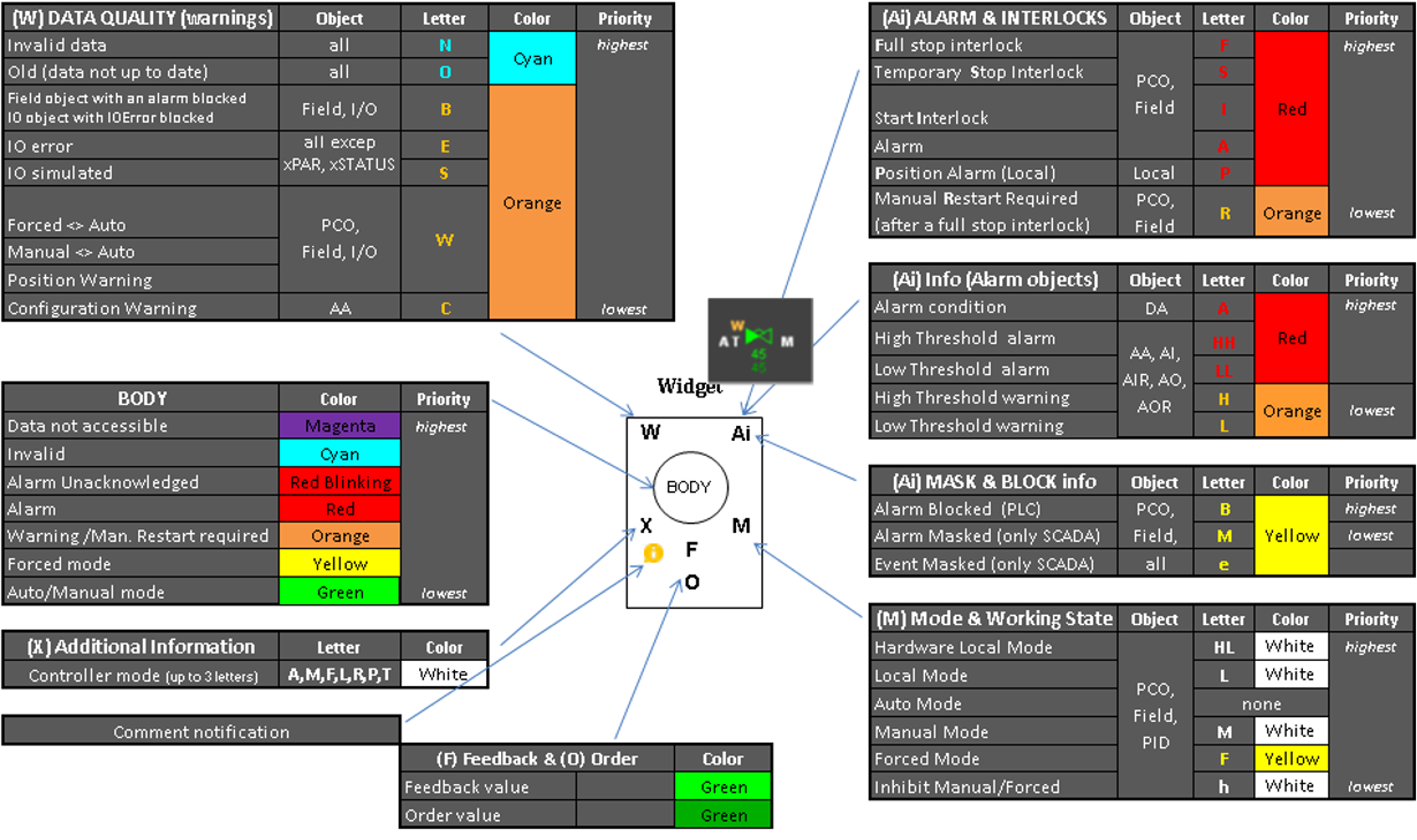}
\caption{An image of the documentation provided to UNICOS operators to decode a widget \cite{unicoscpcwidgethelp}}
\label{fig:unicos_widget_documentation}
\end{figure}

\noindent Another pain point for UNICOS operators relates to the manual and time-consuming effort to diagnose the root cause of an issue. When a widget indicates an error or alarm, the root cause of the issue would need to be identified. It is important to note that, in UNICOS, devices are represented at a very atomic level. As such, a widget might not necessarily represent a device, but rather, a widget may represent a single signal coming from a device, with multiple widgets possibly representing multiple signals coming from a single device. This effectively results in a hierarchy of widgets that affect each other significantly. Therefore, root cause analysis would usually be carried out by traversing such a hierarchy of widgets - essentially inspecting widgets associated with the widget in question, including its master, parents, and children, recursively tracing the issue to its source. The window in Figure~\ref{fig:unicos_widget_hierarchy} below shows the master, parents, and children of a certain widget in the UNICOS user interface. Root cause analysis is usually carried out by an operator, but since this process requires manual effort and is time-consuming, this work aims to leverage AI to automate root cause analysis.\\

\begin{figure}[H]
\centering
\includegraphics[width=0.7\textwidth]{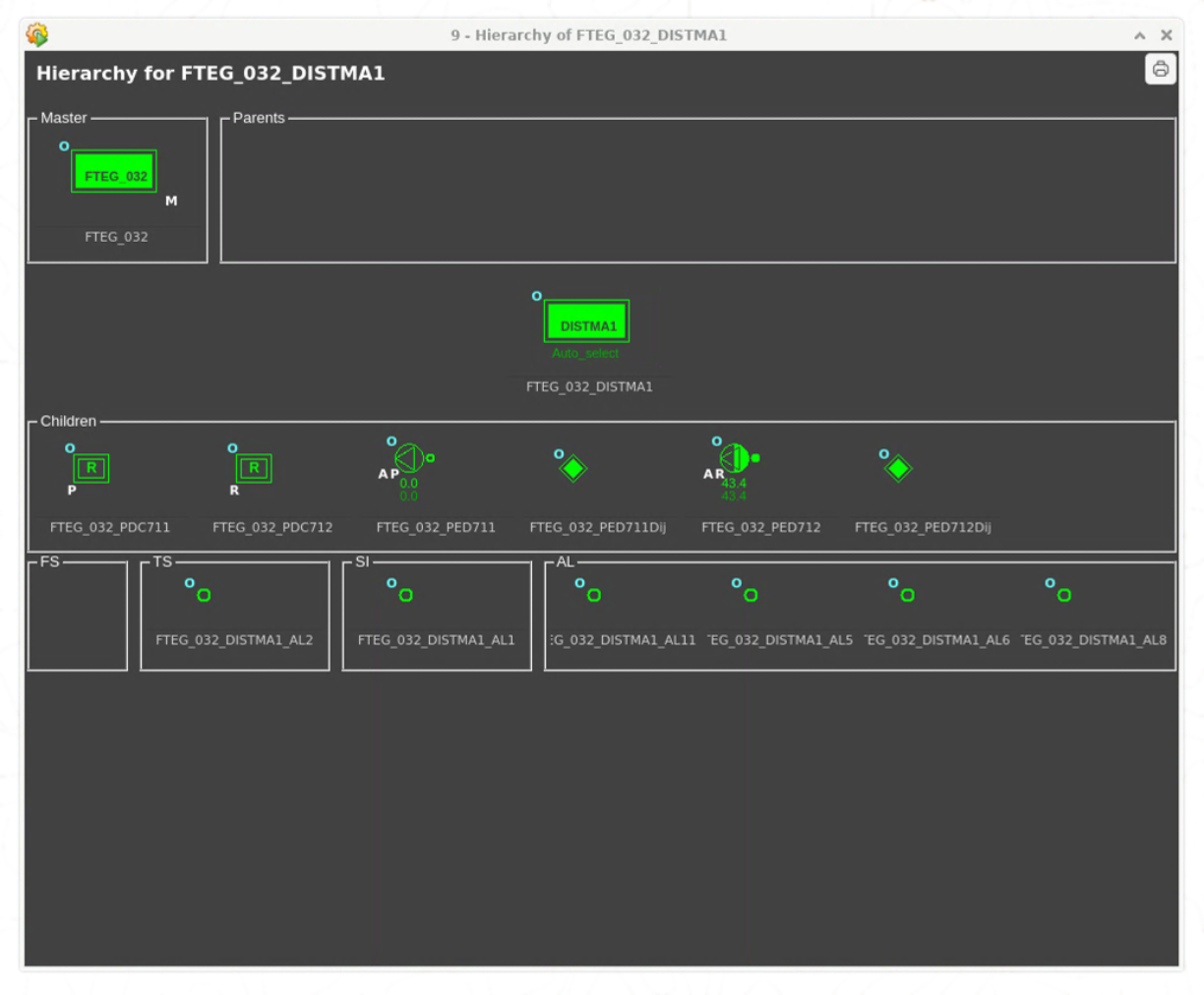}
\caption{A screenshot showing the hierarchy of a widget in the UNICOS user interface}
\label{fig:unicos_widget_hierarchy}
\end{figure}

\noindent UNICOS maintainers also face a pain point when debugging and determining exactly which datapoint elements (DPEs) are responsible for animating different parts of a widget. Signals from devices in the form of DPEs are each responsible for animating different parts of a widget. Logic, expressed as conditional statements in code, sits between such signals and the widgets, thereby determining how each widget is to be animated. When widgets are not animated as expected or exhibit anomalous behaviour, debugging must be carried out by maintainers to investigate and determine exactly which DPEs are responsible for animating each part of the widget. However, the animation logic has become difficult to trace and understand owing to the complex codebase structure and `spaghetti code', leading to this manual process taking maintainers more than 30 minutes on average. As a result, this work streamlines the tracing of DPEs responsible for animating different parts of the widget by employing AI to automate this process. 

\subsection{Additional Benefits}
In addition to addressing each of the pain points mentioned, this project is likely to come with additional benefits, including experience with multi-modal models, experimentation with retrieval augmented generation (RAG) in a complex context, creation of reusable architecture, and foundation for human-in-the-loop (HMI) interfaces.\\

\noindent Due to the visual nature of the UNICOS interface and widgets, it is evident that computer vision (CV) models and large language models (LLMs) supporting image inputs will be used throughout the project. As such, the completion of this project is expected to offer a wealth of hands-on experience with multi-modal models that can take both text and images as input, a key capability for future AI applications in control systems.\\

\noindent Furthermore, this project presents an opportunity to experiment with RAG in a complex context. At the time of writing, there have been no prior recorded attempts at integrating RAG with UNICOS. As such, this project would provide valuable insight into the performance and feasibility of implementing RAG in real-world UNICOS applications, and in a wider sense, industrial control applications. Given the opportunity to tackle challenges like partial documentation and complex codebases, the results produced by this project would provide a clear signal into whether RAG has a promising future in such use cases and contexts.\\

\noindent The completion of this project would also produce reusable architecture. The modular architecture of the finished project could be reused or extended for other HMI features and expert assistance tools. Components in this modular architecture include the AI assistant's user interface (UI), the UNICOS REST API endpoints, the multi-agent system, abstracted database modules, and more.\\

\noindent Finally, this project could act as a foundation for HMI interfaces as it explores a new interaction model between UNICOS operators and the system. As such, it could serve as a prototype for future assistive HMI designs.

\break

\section{METHODOLOGY}

\subsection{System Overview}

\begin{figure}[H]
\centering
\includegraphics[width=1\textwidth]{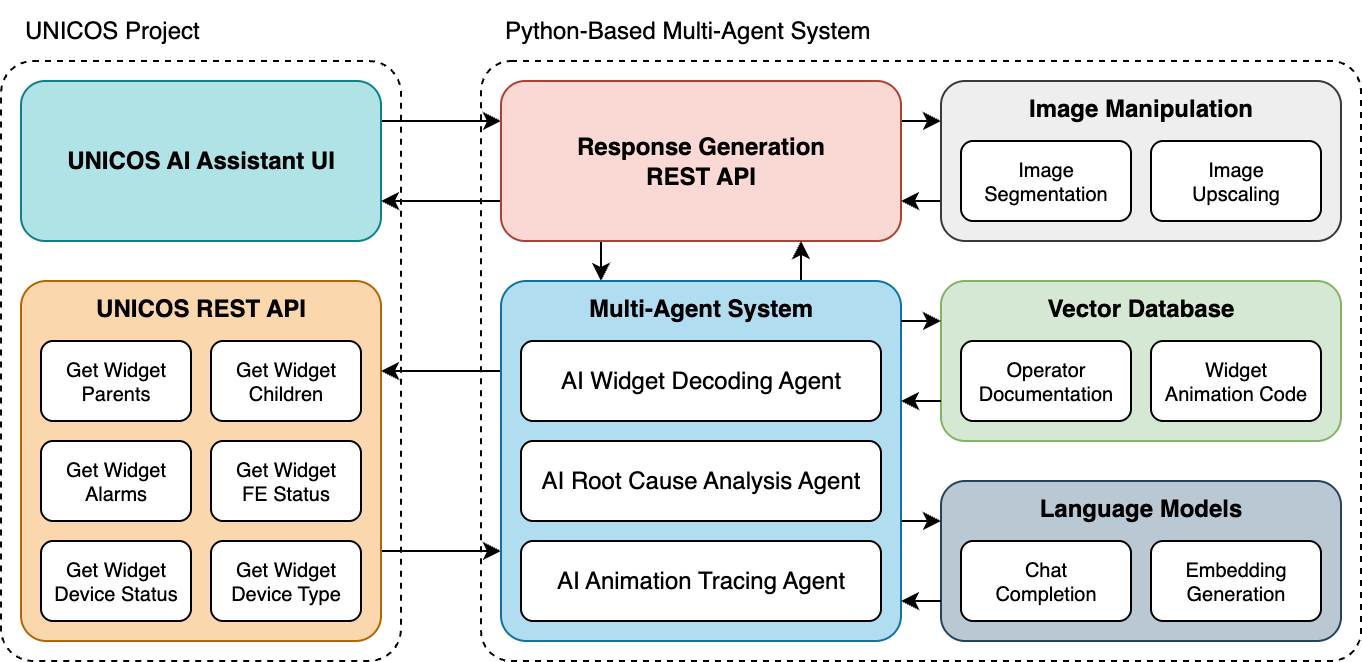}
\caption{A system diagram providing a visual overview of the system developed}
\label{fig:system_diagram}
\end{figure}

Figure~\ref{fig:system_diagram} above provides a visual overview of the various components in the system developed. Overall, the system consists of two major parts. The first is the UNICOS side of the system, which is essentially an extension of existing UNICOS components using CTRL code. The new AI Assistant UI is implemented in this part of the system, along with the REST API endpoints that would be called by the agent performing root cause analysis to retrieve real-time widget information. This part is to be installed as a package to an existing project.\\

\noindent The second part is the multi-agent system, which is a separate part of the system developed in Python and hosted on a regular virtual machine (VM). This part of the system exposes REST API endpoints for response generation when a user sends a query through the AI Assistant UI. It also hosts the entire logic behind the response generation, which includes 1 supervisor agent and 3 specialised worker agents. The supervisor agent delegates tasks to the most appropriate specialised worker agent, whereas the specialised worker agents comprise an AI Widget Decoding Agent, an AI Root Cause Analysis Agent, and an AI Widget Part Animation DPE Tracing Agent.\\

\noindent The second part of the system also includes the vector database, which stores the documentation usually available to UNICOS operators for the purposes of decoding a widget and performing root cause analysis. The vector database will be queried by the AI Widget Decoding Agent and AI Root Cause Analysis agent during RAG, thereby providing agents with the same materials available to human operators when carrying out their duties. The vector database also stores the code related to the animation of widgets in a separate collection. Its contents will be queried by the AI Widget Part Animation DPE Tracing Agent in order to jump through the code and trace the DPEs, providing it with the very resources UNICOS maintainers would have access to when debugging and determining which DPEs are responsible for animating widgets.\\

\noindent Language model modules also form part of the second part of the system, consisting of a chat completion module and an embedding generation module. These not only form the basis of the multi-agent system, but are highly abstracted and modular such that they support the easy swapping of underlying models to be used with the system, and are easily extensible for any models released in the future to a point where it is plug-and-play. Image manipulation modules also make up the second part of the system, handling image segmentation and upscaling tasks in the preprocessing stage for each user query to ensure that each widget image is properly formatted and visible before further analysis is carried out. 

\subsection{AI Widget Decoding}

\subsubsection{Image Segmentation}
Unlike for panels, UNICOS does not support directly taking a screenshot of an individual widget - this is because widgets are not panels, but rather elements within a panel. Given that UNICOS does have built-in support for taking a screenshot of the entire panel, this was decided as a starting point to segment the image. Given screenshots of entire panels, the initial thought was to train an object detection machine learning model to recognise widgets and subsequently extract them individually. However, this approach was found to require too much training data and annotation work, requiring a significant investment in human labour. In order to better focus on other parts of the project, it was decided that another approach would be optimal if it exists.\\

\noindent It was then discovered that each widget within a panel comes with a distinct solid white boundary around it, with a significant contrast against a grey background. It was also discovered that a workaround could be used to temporarily hide all elements in a panel except the widget in question before taking a screenshot of the entire panel. Such discoveries were taken advantage of and the OpenCV2 Python library was used to segment the widget from the panel screenshot. The specific steps are as follows:
\begin{enumerate}
    \item Hide all elements except the desired widget temporarily and take a screenshot of the entire panel
    \item Convert the screenshot to greyscale
    \item Use canny edge detection to easily detect the distinct solid white boundary around the widget in the screenshot
    \item Extract all contents of the image within detected boundary
\end{enumerate}

\subsubsection{Image Upscaling}
A significant problem was encountered during data extraction where the LLM supporting image input was wrongly extracting data due to the low resolution of the widget image. This was especially the case when symbols in a widget overlapped with each other. As a solution, an open-source Enhanced Deep Residual Networks (EDSR) model was used to upscale each widget image by a factor of 4. This process was carried out after image segmentation and before data extraction, and solved the issue of inaccurate data being extracted. It should be noted that the EDSR model is extremely lightweight, fitting completely on a CPU setup with no GPU required. The EDSR model is available through the super-image Python library, making its integration into the existing code quick and hassle-free.

\subsubsection{Data Extraction}
For data extraction, the upscaled image of the widget from the previous step would be passed into an LLM that supports image input. Specifically, the llama-4-maverick-17b-128e-instruct model was chosen for this step due to its support for image input and high performance on benchmarks, achieving greater results than GPT-4o and Gemini 2.0 Flash on a wide range of major benchmarks \cite{nvidia2025llama4}. A low temperature setting with the value set to 0 is put in place to avoid hallucinations, and also since no creativity is required at all for this task.\\

\noindent Prompt engineering techniques were employed to guide and assist the model in carrying out data extraction from the widget. Specifically, few-shot prompting was used to provide 4 example widgets along with all of their corresponding data (top left symbol, top right symbol, etc.) labeled in XML format, effectively creating 4 data-label pairs that the model may learn from and reference when extracting the data. XML was selected as the format of choice because of its structure, namely its clear opening and closing tags, in contrast to JSON which relies on unlabeled and unclear closing brackets, which would create a need for counting brackets. Formatting in such a way ensures that model responses stay consistent with the structure used in the few-shot examples, making the extracted data easily parseable. The details for such an implementation are as follows:
\begin{enumerate}
    \item Insert the 4 example widget-label pairs as user-assistant message pairs into a message list
    \item Append the actual widget to be queried with its data to be extracted as a user message to the message list
    \item When the message list is sent to the model that supports image input, a response will be returned with the actual widget's extracted data in the same consistent XML format as the examples
\end{enumerate}

\subsubsection{Documentation Retrieval}
For documentation retrieval, a vector database is used to store parts of the documentation helpful to the agents for carrying out tasks, along with their respective vector embeddings. This is to allow searches on the documentation to be carried out semantically for the most relevant sections to a certain query, as opposed to keyword search which would look for exact word or phrase matches.\\

\noindent MongoDB was chosen as the vector database to be used because of its document model, allowing each part of the documentation and corresponding vector embeddings to be stored as key-value pairs in documents. In the application, the following structure was used for each document in the MongoDB database, with each document representing a part of the documentation:

\begin{table}[H]
    \centering
    \begin{tabular}{|p{0.125\textwidth}|p{0.65\textwidth}|}
         \hline
         \textbf{Key} & \textbf{Value} \\
         \hline
         data & string containing unstructured text representing the actual substance of the documentation \\
         \hline
         embeddings & vector embedding that encodes the string containing the unstructured text \\
         \hline
    \end{tabular}
    \caption{The key-value pairs making up each MongoDB document representing a part of the documentation}
    \label{tab:documentation_kv_pairs}
\end{table}

\noindent The vector embeddings stored in each document are used as a search index during semantic search to locate the document if its contents are semantically similar to the search query.\\

\noindent In contrast to a significant number of other systems in the industry, chunking is not carried out by tokens for document splitting. Instead, chunking is carried out by page or topic of documentation owing to the relatively small size of each page and topic supporting this approach. Effectively, most pages within the documentation are a few paragraphs long and doing so avoids the issue of losing context seen in many other systems, where even leaving a buffer for overlap between chunks would not preserve enough context.\\

\noindent An abstracted layer above MongoDB is implemented in the system as a vector database module. This allows the underlying vector database to easily be swapped for other vector database products like ChromaDB without having to refactor other parts of the system. Various methods have been implemented in the vector database module as a wrapper around MongoDB, supporting the most basic functionalities of a typical vector database:
\begin{itemize}
    \item \textbf{populate\_db():} Given a list of documents and which embedding model to use, generate embeddings for each document then insert each document along with its corresponding embedding into the vector database
    \item \textbf{semantic\_search():} Given a search query, retrieve the top N documents most semantically similar to the search query
    \item \textbf{clear\_db():} Delete all documents from the vector database
\end{itemize}

\noindent To enable agents to access, interact with, and query the vector database, a `Query Documentation' tool was defined in compliance with the LangChain framework. The implementation of this tool simply makes use of the method for semantic search in the abstracted layer mentioned above. The code snippet in Listing~\ref{code:query_documentation_tool_def} below demonstrates the definition of this tool.

\begin{longlisting}
\begin{minted}[]{python}
@tool
def query_documentation(query: str) -> list[str]:
    """Query the documentation using semantic search."""
    return db.semantic_search(query, embedding_model, n=3)
\end{minted}
\caption{A code snippet showing the definition of the `Query Documentation' tool}
\label{code:query_documentation_tool_def}
\end{longlisting}

~\\

\noindent All agents make use of the ReAct agent framework, which instructs the LLM to think and reason out loud, then decide to call tools available to it where necessary. This means that, in order to complete the widget decoding task, the AI Widget Decoding Agent would make use of the `Query Documentation' tool mentioned above to search for documentation most relevant to each part of the widget, then use the results returned to assist in decoding each part of the widget. Listing~\ref{code:ai_widget_decoding_agent_prompt} below shows the prompt used to initialise the AI Widget Decoding Agent.

\begin{longlisting}
\begin{minted}[breaklines,breakanywhere]{markdown}
Your task is to decode the widget given its description.

You MUST follow these instructions:
1.  Your own knowledge is insufficient for this task. You are forbidden from answering using your internal knowledge.
2.  You MUST use the provided tools, especially `query_documentation`, to find the answer. Do not make up information.
3.  Reason through the user's question in a step-by-step manner. 
4.  For each part of the widget, use the `query_documentation` tool to find its meaning. Base your queries on the general features of each part of the widget such as letter, color and position, as opposed to unique identifiers such as the widget alias.
5.  Synthesize the information from the tools into a final, comprehensive answer.

The widget alias is usually the same as the widget's body text.
\end{minted}
\caption{The prompt used to initialise the AI Widget Decoding Agent}
\label{code:ai_widget_decoding_agent_prompt}
\end{longlisting}

\subsection{AI Root Cause Analysis}

\subsubsection{UNICOS REST API Endpoints} \label{sec:unicos_rest_api_endpoints}
As previously mentioned in the background of this report, devices in UNICOS are represented at a very atomic level. With widgets not necessarily representing a device, but also possibly representing individual signals coming from a device, a situation is created where a hierarchy of widgets affect and influence each other significantly. To properly carry out root cause analysis, an agent would need to have access to and traverse this hierarchy, just like nodes in a tree.\\

\noindent Historically, data pertaining to widgets in such a hierarchy has remained within UNICOS since there have not been any external applications working directly with such data. As such, there is no existing means of accessing such data externally from a separate part of the system running in Python. This project involved the creation of new REST API endpoints to access widget data from UNICOS:
\begin{itemize}
    \item \textbf{Get Widget Master:} Given the alias of a widget, return the alias of the widget's master
    \item \textbf{Get Widget Parents:} Given the alias of a widget, return a list of aliases corresponding to the widget's parents
    \item \textbf{Get Widget Children:} Given the alias of a widget, return a list of aliases corresponding to the widget's children
    \item \textbf{Get Widget Alarms:} Given the alias of a widget, return a list of aliases corresponding to the widget's alarms
    \item \textbf{Get Widget Frontend Status:} Given the alias of a widget, return the frontend status code of the widget, which indicates the connection status of the associated device
    \item \textbf{Get Widget Device Status:} Given the alias of a widget, return the device status code of the widget, which encodes the operational status of the associated device
    \item \textbf{Get Widget Device Type:} Given the alias of a widget, return the device type of the associated device
\end{itemize}

\noindent Each of the endpoints above was designed to be running in the background in UNICOS, with the background service spawning automatically upon startup of the project. By exposing these endpoints externally, necessary data about each widget may be retrieved and the hierarchy may be traversed by the agent to perform root cause analysis. Similar to the definition of the `Query Documentation' tool in Listing~\ref{code:query_documentation_tool_def} above, each endpoint is wrapped into its corresponding tool according to the LangChain framework to enable agents' interaction with the endpoints through the tools. Furthermore, the AI Root Cause Analysis Agent also makes use of the ReAct agent framework, meaning that it will think and reason out loud, then make use of the tools defined to traverse the hierarchy and find the problem. Listing~\ref{code:ai_root_cause_analysis_agent_prompt} below shows the prompt used to initialise the AI Root Cause Analysis Agent.

\begin{longlisting}
\begin{minted}[breaklines,breakanywhere]{markdown}
Your task is to determine the root cause of the widget issue.

You MUST follow these instructions:
1.  Your own knowledge is insufficient for this task. You are forbidden from answering using your internal knowledge.
2.  You MUST use the provided tools, especially `query_documentation`, to find the answer. Do not make up information.
3.  Reason through your task in a step-by-step manner.
4.  To find the root cause of the issue, you must first decode the meaning of this widget. This might already have been done by the widget decoding agent. 
5.  Then, traverse the hierarchy, finding the device's parents, grandparents, children, and grandchildren, if any, and see if the problem stems from any of them.
6.  For each device, you can get the frontend status code, which tells you info about connection, and you can also get the device status code, which tells you info about the device itself. They are encoded in special ways so you need to query the documentation to understand the codes and come to a conclusion. Frontend status code is universal but device status code is specific to each device, so you also need to use the available tools to get the device type before finding the right documentation to decode the device status code for that specific device.
7.  Synthesize the information from the tools into a final, comprehensive answer.

The widget alias is usually the same as the widget's body text.
\end{minted}
\caption{The prompt used to initialise the AI Root Cause Analysis Agent}
\label{code:ai_root_cause_analysis_agent_prompt}
\end{longlisting}

\subsubsection{Documentation Retrieval}
During AI Root Cause Analysis, documentation retrieval is carried out in the same manner and from the same source as AI Widget Decoding. This is because the documentation contained within the vector database spans not only widget symbol meanings, but also device status code meanings and frontend status code meanings, which are used for root cause analysis. As such, both the AI Widget Decoding Agent and AI Root Cause Analysis Agent use and share the same `Query Documentation' tool.

\subsection{AI Tracing of Widget Part Animation DPEs}

\subsubsection{Code Retrieval} \label{sec:code_retrieval}
In order for the agent to trace exactly which DPEs are responsible for animating different parts of a widget, access to the code related to the animation of widgets is required. To achieve this, all such code was loaded into a vector database. Given that MongoDB was already used as a vector database to store documentation for AI Widget Decoding and AI Root Cause Analysis, the code was simply stored in a separate collection within the same MongoDB database, thereby also being represented in a document model. The following structure was used for each document in the collection, with each document representing a single method in the codebase:

\begin{table}[H]
    \centering
    \begin{tabular}{|p{0.155\textwidth}|p{0.65\textwidth}|}
         \hline
         \textbf{Key} & \textbf{Value} \\
         \hline
         file\_name & name of the file where the code is located \\
         \hline
         method\_name & name of the method where the code is located \\
         \hline
         data & string containing all the code in the method, including code comments and docstrings \\
         \hline
         embeddings & vector embedding that encodes the string containing the code \\
         \hline
    \end{tabular}
    \caption{The key-value pairs making up each MongoDB document representing a method in the codebase}
    \label{tab:codebase_kv_pairs}
\end{table}

\noindent As seen in Table~\ref{tab:codebase_kv_pairs} above, additional metadata in the form of the code's file and method name is added to each document. Chunking is carried out by method as each method could be treated as a logical unit, further supported by the reasonably small size of each method. An additional class was implemented in the code that extends the abstract vector database module specifically to handle operations related to the code. In addition to semantic search, which is inherited from the base class of the abstract vector database module, the following additional methods were implemented:

\begin{itemize}
    \item \textbf{file\_name\_search():} Given a file name, return the code for all methods located within that file
    \item \textbf{method\_name\_search():} Given a method name, return all the code for that method
    \item \textbf{multi\_filter\_search():} Given a combination of file name, method name, and semantic search query, return the code for all methods that fit all the constraints
\end{itemize}

\noindent Much like the endpoints for AI Root Cause Analysis, each method is wrapped into its corresponding tool in line with the LangChain framework, enabling agents to perform searches on the codebase by file name, method name, semantic search, or a combination thereof. The ReAct agent framework is also used here, meaning that the agent will think and reason out loud before invoking the tools to navigate the codebase and trace the DPEs responsible for animating each part of the widget. Listing~\ref{code:ai_widget_part_animation_dpe_tracing_agent_prompt} below shows the prompt used to initialise the AI Widget Part Animation DPE Tracing Agent. 

\begin{longlisting}
\begin{minted}[breaklines,breakanywhere]{markdown}
Your task is to trace which datapoint elements are responsible for animating the widget, given a description of the widget's appearance and the widget's alias.

You MUST follow these instructions:
1.  Your own knowledge is insufficient for this task. You are forbidden from answering using your internal knowledge.
2.  You MUST use the provided tools, especially `query_codebase` and `query_codebase_by_method`, to find the answer. Do not make up information.
3.  Reason through the user's question in a step-by-step manner.
4.  To find the datapoint element that animates the widget, you must make use of the available tools to find the device type and explore the codebase for code that animates widgets of that device type, finally tracing the datapoint element names that are responsible for animating the specific widget.
5.  Queries to the codebase may be conducted by semantic search or by exact method name, depending on which is more appropriate for the specific query at that point in time.
6.  For example, tracing for a device of type AnalogDigital (ANADIG) would involve the following hops: unSimpleAnimation_WidgetConnect() -> CPC_AnaDig_WidgetDPEs() -> unSimpleAnimation_WidgetAnimationCB() -> CPC_AnaDig_WidgetAnimation() -> cpcGenericObject_WidgetAnimationDoubleStsReg()
7.  As another example, tracing for a device of type ProcessControlObject (PCO) would involve the following hops: unSimpleAnimation_WidgetConnect() -> CPC_ProcessControlObject_WidgetDPEs() -> unSimpleAnimation_WidgetAnimationCB() -> CPC_ProcessControlObject_WidgetAnimation() -> cpcGenericObject_WidgetAnimationDoubleStsReg()
8.  Always remember the order is as follows: unSimpleAnimation_WidgetConnect() -> <DeviceType>_WidgetDPEs() -> unSimpleAnimation_WidgetAnimationCB() -> <DeviceType>_WidgetAnimation() -> cpcGenericObject_WidgetAnimationDoubleStsReg()
9.  However, a surface-level trace is not enough, each relevant method must be thoroughly analysed in order to determine which datapoint element names are responsible for animating the specific widget, paying special attention to the exact order of parameters when they are being passed into methods.
10. For each datapoint element name, explain exactly which part of the widget it animates and what effect it produces. Use conditional statements from the codebase to justify your answer.

The widget alias is usually the same as the widget's body text.
\end{minted}
\caption{The prompt used to initialise the AI Widget Part Animation DPE Tracing Agent}
\label{code:ai_widget_part_animation_dpe_tracing_agent_prompt}
\end{longlisting}

\break

\section{RESULTS AND DISCUSSION}

\subsection{AI Widget Decoding}
\begin{figure}[H]
\centering
\includegraphics[width=1\textwidth]{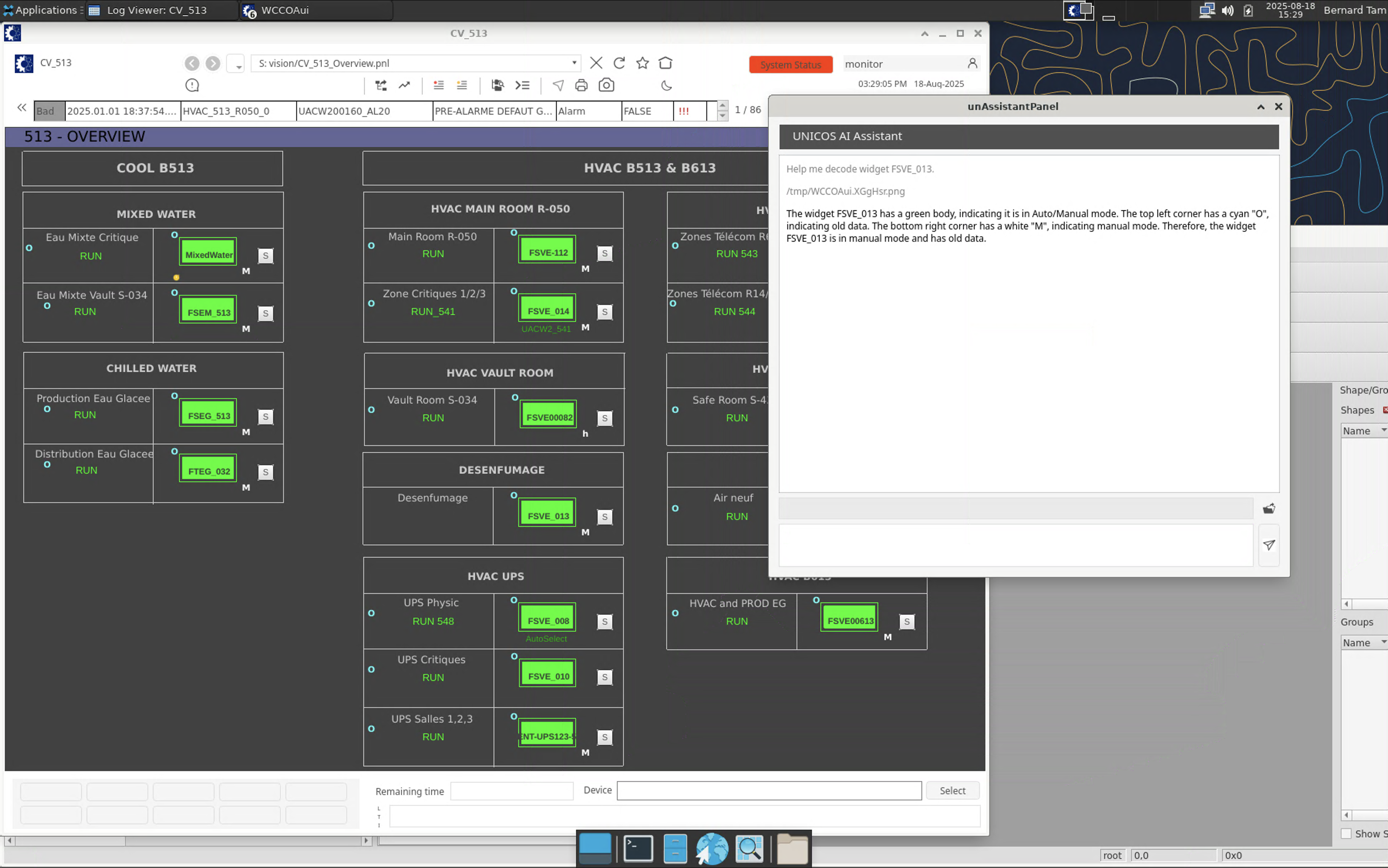}
\caption{A screenshot demonstrating AI Widget Decoding}
\label{fig:demo1}
\end{figure}

\noindent The screenshot in Figure~\ref{fig:demo1} above demonstrates AI Widget Decoding in action. The AI Assistant window, which was developed in this project as a new addition to UNICOS, is shown in the foreground. As seen within the window, the conversation history between the user and the system is shown, with the user sending a message requesting help in decoding the widget `FSVE\_013', and attaching the said widget. In response to the user's query, the system is shown to respond with information about the widget in its decoded form.\\

\noindent The system's response suggests that the widget's green body indicates that it is in Auto/Manual mode. It also suggests that the cyan `O' in the top left corner indicates old data. The system finishes with stating that the white `M' in the bottom right corner confirms that it is in manual mode.\\

\noindent The descriptions produced by the system fully align with information found in the documentation provided to UNICOS operators, as seen in Figure~\ref{fig:unicos_widget_documentation}. This successfully demonstrates the system's ability to accurately decode the widget, correctly explaining the meaning of all 3 elements in the widget. 

\subsection{AI Root Cause Analysis}
\begin{figure}[H]
\centering
\includegraphics[width=1\textwidth]{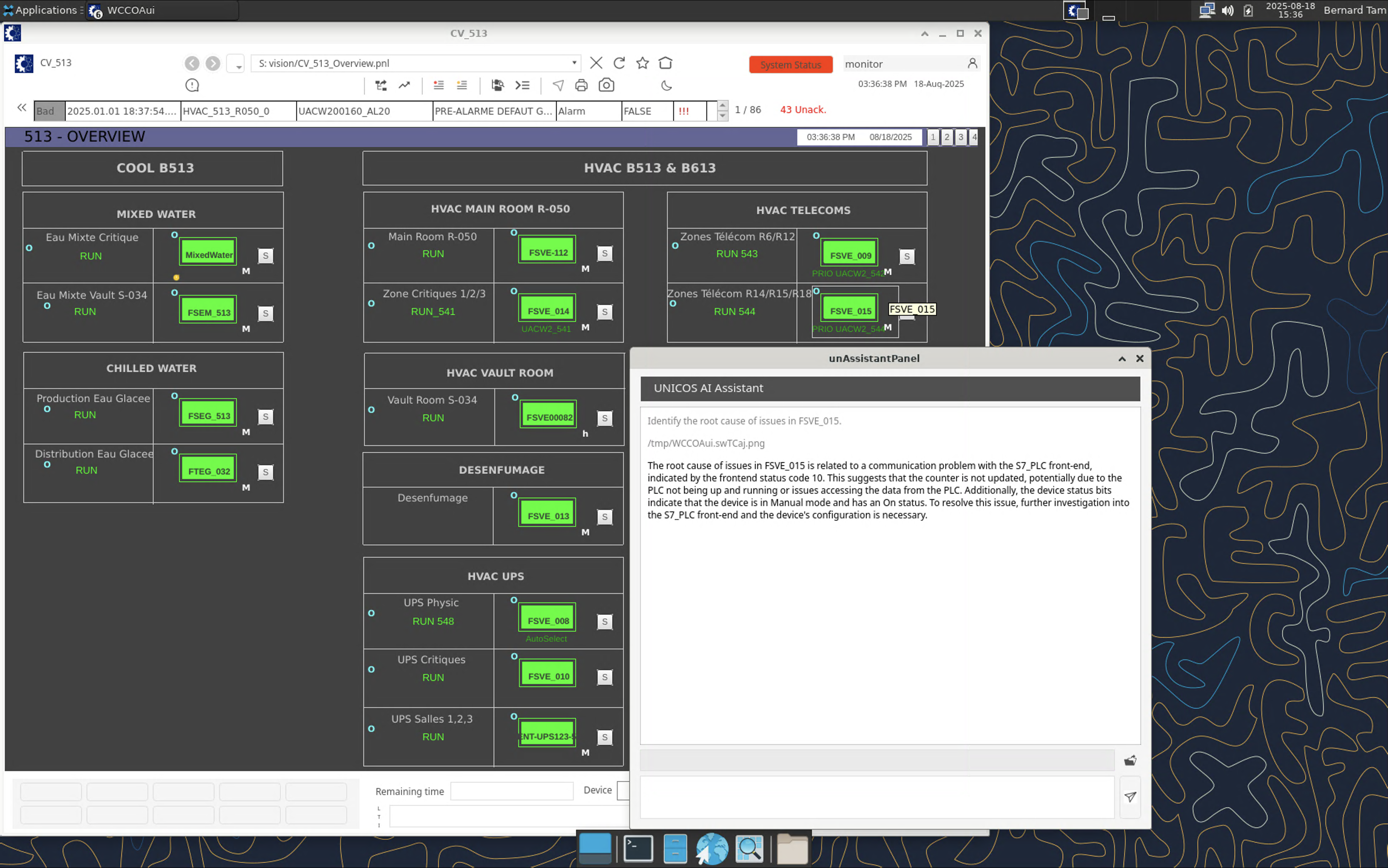}
\caption{A screenshot demonstrating AI Root Cause Analysis}
\label{fig:demo2}
\end{figure}

\noindent The screenshot in Figure~\ref{fig:demo2} above demonstrates AI Root Cause Analysis in action. Similar to Figure~\ref{fig:demo1} that demonstrates AI Widget Decoding, the AI Assistant window is shown in the foreground. This same window is used by users to interact with the system for AI Root Cause Analysis. The window shows a different conversation history between the user and the system where the user makes a request to identify the root cause of issues with the widget `FSVE\_015' and attaches the said widget. The system is shown to have responded to the user's query with an explanation of the potential root cause and suggestions to resolve the issue.\\

\noindent In the system's response, the root cause is suggested to be a communication problem with the frontend, referencing the frontend status code 10 and that such a code indicates a certain counter is not updated. The system's response also states that the device status bits point to the device being on and running in manual mode. It finishes with suggesting a further investigation into the frontend and configuration of the device.\\

\noindent In order to be aware that the frontend status code was 10, the system must have been able to access the data through the `Get Widget Frontend Status' tool mentioned in Section \ref{sec:unicos_rest_api_endpoints} above, then access the documentation about frontend status codes to determine exactly what a value of 10 meant. Similarly, to retrieve the device status bits, the system must have been able to access the `Get Widget Device Status' tool also mentioned in Section \ref{sec:unicos_rest_api_endpoints} above and subsequently access documentation about what each device status bit maps to. This successfully demonstrates the system's ability to access live data from devices and understand documentation to make sense of the live data retrieved. The system's output also highlights its ability to make sensible suggestions on how to proceed in investigating the issue. 

\subsection{AI Tracing of Widget Part Animation DPEs}
\begin{figure}[H]
\centering
\includegraphics[width=1\textwidth]{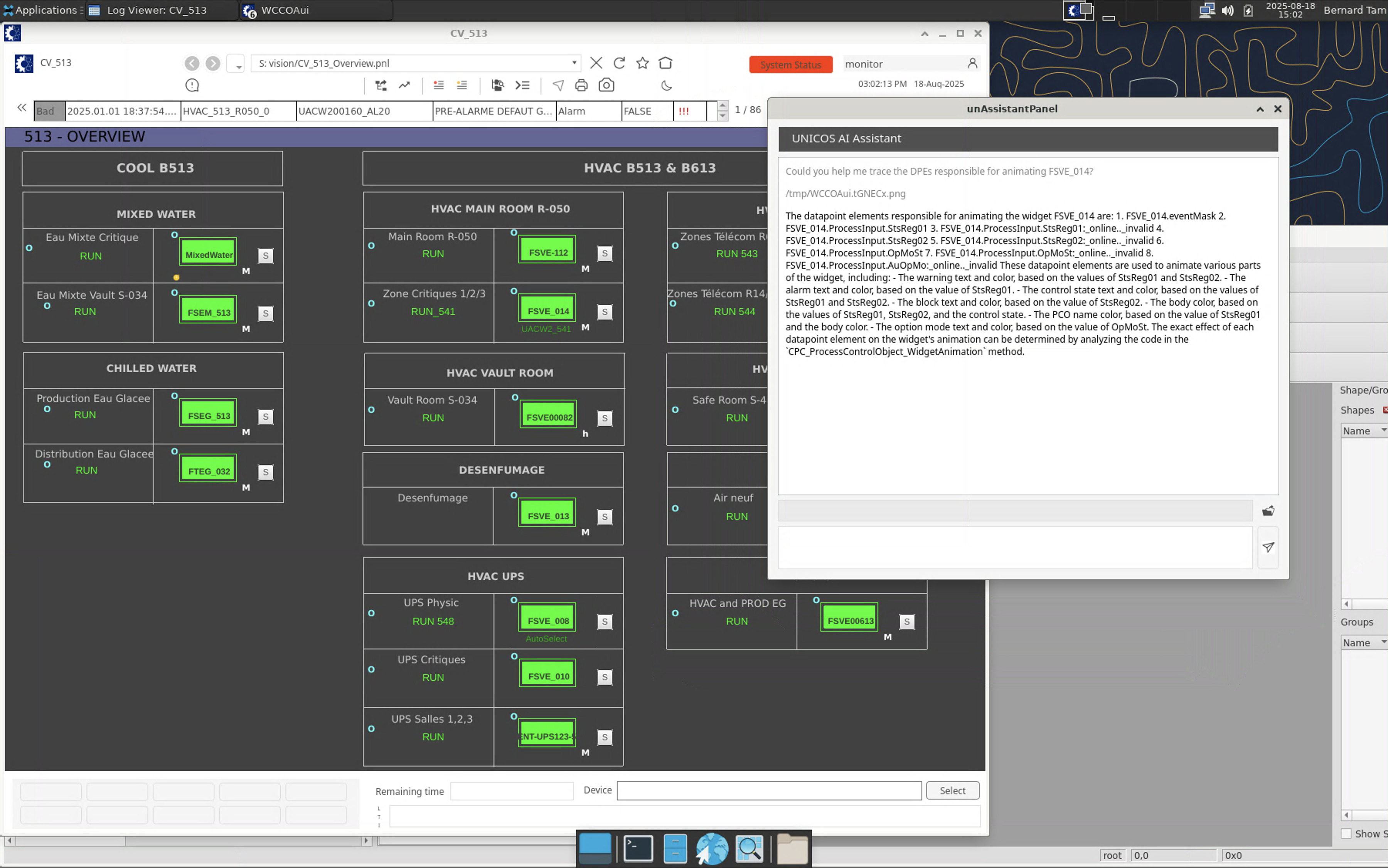}
\caption{A screenshot demonstrating AI Tracing of Widget Part Animation DPEs}
\label{fig:demo3}
\end{figure}

\noindent The screenshot in Figure~\ref{fig:demo3} above demonstrates AI Tracing of Widget Part Animation DPEs in action. This use case also makes use of the same AI Assistant window for users to interact with the system, just like both the previous use cases. As such, the same window is shown in the foreground. In the conversation history within the window, the user requests the system to assist in tracing the DPEs responsible for animating the widget `FSVE\_014' and attaches the said widget. The system then responds with the DPEs responsible for animating the widget and the method name containing the relevant code.\\

\noindent The DPEs responsible for animating the widget are returned in the format of a numbered list, along with an explanation of which parts of the widget each DPE animates, including the warning text, body colour, and more. The response ends with the system directing the user towards a certain method in the code to further explore the exact effect of each DPE on the widget's animation.\\

\noindent To successfully trace the DPEs responsible for animating the widget, the system would have had to make use of the tools mentioned in Section \ref{sec:code_retrieval} above to retrieve methods and sections of code by means of semantic search, file name search, method name search, or a combination thereof. By using such tools to navigate the codebase, jump between different methods, and trace the DPEs responsible for animating the widget, the system demonstrates its ability to find its way around complex codebase structures, understand what each section of code does, and purpose these skills towards a specific use case. The system's response further illustrates its ability to explain what it has understood about the code to the user and point them in the right direction should they wish to dive deeper into the code themselves.

\break

\section{CONCLUSION}
In this project, a multi-agent system was successfully developed to address three major pain points experienced by UNICOS operators and maintainers: widget decoding, root cause analysis, and the tracing of DPEs responsible for animating widgets. As demonstrated by the results of this project, the system establishes the capability of AI in substantially reducing cognitive load for operators, speeding up diagnostic processes, and supporting maintainers in navigating a complex codebase structure.\\

\noindent Beyond tackling these immediate challenges, this work represents an important first step in providing AI-powered assistance to operators in the CCC. By incorporating multi-modal reasoning, retrieval augmented generation, and a modular architecture, the system establishes both a practical proof-of-concept and a methodological foundation for future research. Beyond its immediate benefits, this work also provides a basis for subsequent exploration into expanding the scope of intelligent support functionalities within UNICOS.

\section{FUTURE WORK}
In terms of future work, an area for exploration could be the fine-tuning of models with CERN-specific data. In doing so, models would be trained to better comprehend CERN-specific vocabulary and terminology that would not otherwise be understood by foundation models. When such a fine-tuned model is used as the reasoning model powering the multi-agent system, it is expected to yield responses of a higher quality.\\

\noindent Another possible direction of future work would be extending usage of the vector database storing the codebase to more contexts and purposes. The vector database has been populated with code in a way that takes extensibility and multiple use cases into account. As such, it would be sensible to take advantage of the already implemented functionalities that support searching the codebase by file name, method name, semantic similarity, or a combination thereof to extend beyond the tracing of widget animation DPEs.\\

\noindent It would also be helpful to integrate CI/CD pipelines to update the vector database storing the codebase. This is because in case of any change to the code used in production for animating widgets, the updated code must automatically be synchronised into the vector database to reflect the latest changes. This would ensure that responses from the system are always up to date and reflect the current reality. 

\break

% Use \section*{REFERENCES} to avoid REFERENCES being shown in the table of contents
\section{REFERENCES}
\printbibliography[heading=none]

@article{gayet2005unicos,
  title={UNICOS a framework to build industry like control systems: Principles \& Methodology},
  author={Gayet, Philippe and Barillere, Renaud and others},
  journal={10th ICALEPCS},
  year={2005}
}

@article{goralczyk2022cern,
  title={CERN SCADA Systems 2020 Large Upgrade Campaign Retrospective},
  author={Goralczyk, Lukasz and Kostopoulos, Alexandros Foivos and Tournier, Jean-Charles and Schofield, Brad},
  journal={JACoW},
  pages={156--160},
  year={2022}
}

@article{silvola2022devops,
  title={DevOps and CI/CD for WinCC Open Architecture Applications and Frameworks},
  author={Silvola, Riku-Pekka and Sargsyan, Laura},
  journal={JACoW},
  pages={281--285},
  year={2022}
}

@online{introductiontoctrl,
  author       = {{ETM professional control GmbH}},
  title        = "Control programming - Introduction to CTRL",
  url = "https://www.winccoa.com/documentation/WinCCOA/3.18/en_US/Control_Grundlagen/Control_Grundlagen-15.html",
  urldate = {2025}
}

@online{projectdescription,
  author       = "Jean-Charles Tournier",
  title        = "UNICOS AI Operator Assistant",
  url = "https://codimd.web.cern.ch/kbBPE8mjQBKd9tBmMJd22Q",
  urldate = {2025-06}
}

@online{unicoscpcwidgethelp,
  author       = {{European Organization for Nuclear Research}},
  title        = "UNICOS CPC Widget Help",
  url = "https://unicos.web.cern.ch/pages/documentation.html",
  urldate = {2025}
}

@online{nvidia2025llama4,
  author       = {{NVIDIA Corporation}},
  title        = "Llama 4 Models — NVIDIA NeMo Framework User Guide",
  url = "https://docs.nvidia.com/nemo-framework/user-guide/latest/vlms/llama4.html",
  urldate = {2025}
}

\end{document}